# Noncommutative Space-Time in DSR theories

**Clemens Heuson**

Zugspitzstr. 4, D-87493 Lauben, Germany
email: clemens.heuson@freenet.de

Space-time coordinates in DSR theories with two invariant scales based on a dispersion relation with an energy independent speed of light are introduced by the demand, that boost and rotation generators are invariant under a transformation from SR to DSR variables. This turns out to be equivalent to a recent suggestion postulating the existence of plane wave solutions in DSR theories. The momentum space representation of coordinates is derived, yielding a noncommutative space-time and the deformed algebra.

## 1. Introduction

In the last years there was an increased interest in the so called doubly or deformed special relativity theories (DSR), involving two invariant scales, a velocity scale and a length scale or energy/momentum scale, see the review in [1] and papers therein. An additional invariant fundamental length or energy scale is in clear contradiction to standard Lorentz contraction, so one expects that special relativity (SR) must be changed somehow.

Since most results obtained so far concern the energy-momentum sector, a key open problem of DSR theories is of course the constructon of the proper space-time sector and especially the question of it's noncommutativity, cf. the discussion in [1]. The space-time sector in DSR theories was first investigated in the context of $\kappa$-Poincare algebra in [2],[3] resulting in noncommutative coordinates. More recently in [4] it was suggested, to fix position space by the invariance of $p_\mu x^\mu$ under the nonlinear map or in [5] from a Hamiltonian formalism both yielding coordinates transforming differently from momenta under deformed Lorentz transformations.

In this letter we derive space-time coordinates for a class of DSR theories with energy independent speed of light from the demand that the boost and rotation generators remain invariant under a transformation of SR variables to DSR variables, which turns out to be equivalent to the suggestion in [4]. This provides us with a momentum space representation of noncommuting coordinates, fixing their behaviour under deformed Lorentz transformation and the corresponding deformed algebra.

## 2. Deformed Lorentztransformations

In this section we discuss the properties of deformed Lorentz transformations in general DSR theories with energy independent speed of light. For this aim we first write down some important properties of SR and then the connection with DSR theories. Here and in the following we take as metric $diag(\eta) = (-1, +1, +1, +1)$ and use units $\hbar = c = 1$. We denote the SR linear 4-momenta by $\pi_\mu = (-\epsilon, \boldsymbol{\pi})$ and $\pi^\mu = (\epsilon, \boldsymbol{\pi})$. SR is defined by the invariant dispersion relation

$$\tilde{C} = -\pi^2 = \epsilon^2 - \boldsymbol{\pi}^2 = m^2 \tag{1}$$

The Lorentz transformations for co- and contravariant momenta are from [6]

$$\pi'^{\mu} = \Lambda^{\mu}_{\nu} \pi^{\nu},\ \pi'_{\mu} = \overline{\Lambda}^{\nu}_{\mu} \pi_{\nu},\ \pi^{\mu} = \overline{\Lambda}^{\mu}_{\nu} \pi'^{\nu},\ \pi_{\mu} = \Lambda^{\nu}_{\mu} \pi'_{\nu} \tag{2}$$

where $\Lambda^{\mu}_{\nu}$ denote the standard Lorentz transformations and $\overline{\Lambda}^{\mu}_{\nu}$ the inverse transformations obtained by the replacement $-v \to v$. The Lorentz transformation and its inverse obey the relations $\Lambda^{\mu}_{\rho} \overline{\Lambda}^{\rho}_{\nu} = \delta^{\mu}_{\nu}$ guaranteeing the invariance of the Casimiroperator in (1). The space-time coordinates of special relativity in momentum space are defined by

$$\xi_{\mu} = i\,\frac{\partial}{\partial \pi^{\mu}}\ ,\ \xi^{\mu} = i\,\frac{\partial}{\partial \pi_{\mu}} \tag{3}$$

and consequently the boost and rotation generators are

$$\tilde{M}_{\mu\nu} = \pi_{\nu} \xi_{\mu} - \pi_{\mu} \xi_{\nu} = i \left(\pi_{\nu}\,\frac{\partial}{\partial \pi^{\mu}} - \pi_{\mu}\,\frac{\partial}{\partial \pi^{\nu}}\right) \tag{4}$$

Momenta and coordinates transform as first order tensors under Lorentz transformations, while the boost and rotation generators transform as second order tensors and obey the standard Poincare-Heisenberg algebra with commuting coordinates and the Heisenberg commutation relation between coordinates and momenta. The transformation of derivatives with respect to momenta is obtained from $\frac{\partial}{\partial \pi'^{\mu}} = \frac{\partial \pi^{\nu}}{\partial \pi'^{\mu}}\,\frac{\partial}{\partial \pi^{\nu}}$ as

$$\frac{\partial}{\partial \pi'^{\mu}} = \overline{\Lambda}^{\nu}_{\mu}\,\frac{\partial}{\partial \pi^{\nu}}\ ,\ \frac{\partial}{\partial \pi'_{\mu}} = \Lambda^{\mu}_{\nu}\,\frac{\partial}{\partial \pi_{\nu}} \tag{5}$$

from which one gets the transformation of the SR coordinates $\xi_{\mu}$ and $\xi^{\mu}$ in analogy to (2).

DSR theories involve an second invariant length scale $\ell$, or energy scale $\kappa = 1/\ell$. The connection between the linear SR momenta $\pi$ and the nonlinear DSR momenta $p$ with $p_{\mu} = (-E, \boldsymbol{p})$ and $p^{\mu} = (E, \boldsymbol{p})$ is given by

$$\pi = F(p)\, p\ ,\ p = F^{-1}(\pi)\, \pi \tag{6}$$

with the nonlinear function $F(p, \ell)$ and its inverse $F^{-1}(\pi, \ell)$, where we have interchanged $F$ and $F^{-1}$ compared to the notation in [7], because it is mostly $F$ appearing in our equations. Note that F coincides with the U-map of [4] in momentum space. In DSR theories with variable speed of light $F$ acts differently on energy and momenta $F\, p^{\mu} = (f\, E, g\, \boldsymbol{p})$. In this paper we focus on the simpler case of DSR theories with energy independent speed of light, in which $F\, p^{\mu} = (F\, E, F\, \boldsymbol{p})$. Inserting (6b) in (6a) gives immediately $F^{-1}(\pi) = 1/F(p)$. In the Magueijo Smolin model [8] we have $F = (1 - \ell E)^{-1}$, $F^{-1} = (1 + \ell \epsilon)^{-1}$, while in the model of [9] $F = (1 - \ell^2\, \boldsymbol{p}^2)^{-1/2}$ and $F^{-1} = (1 + \ell^2\, \boldsymbol{\pi}^2)^{-1/2}$. The deformed dispersion relation of DSR theories is now obtained by inserting (6a) into (1)

$$C = -F^2\, p^2 = F^2\, (E^2 - \boldsymbol{p}^2) = \mu^2 \tag{7}$$

The SR momenta $\pi$ transform according the standard linear Lorentz transformation in (2) from which together with (6) the nonlinear deformed Lorenz transformation $L = L(p)$ is derived as $L = F^{-1}\, \Lambda\, F$. Since the deformed Lorentz transformation depends nonlinearly on the momenta, one has to distinguish carefully between the transformation and backtransformation of co- and contravariant momenta.

$$p'^{\mu} = L^{\mu}_{\nu}\, p^{\nu},\ p'_{\mu} = \overline{L}^{\nu}_{\mu}\, p_{\nu},\ p^{\mu} = \overline{L}'^{\mu}_{\nu}\, p'^{\nu},\ p_{\mu} = L'^{\nu}_{\mu}\, p'_{\nu} \tag{8}$$

where $L^{\mu}_{\nu} = A\, \Lambda^{\mu}_{\nu}$, $\overline{L}^{\nu}_{\mu} = A\, \overline{\Lambda}^{\nu}_{\mu}$, $\overline{L}'^{\mu}_{\nu} = A'\, \overline{\Lambda}^{\mu}_{\nu}$, $L'^{\nu}_{\mu} = A'\, \Lambda^{\nu}_{\mu}$. Here $A = A(E, p_1, \ell)$ is a nonlinear function of the momenta $p$. In the model [8] one has $A = (1 - \ell\, E + \ell\, \gamma\, (E - v\, p_1))^{-1}$, while in the model [9] $A = \left(1 - \ell^2\, E^2 + \ell^2\, \gamma^2 (E - v\, p_1)^2\right)^{-1/2}$. The corresponding function $A'$ is obtained by the replacements $E \to E'$, $p_1 \to p'_1$, $v \to -v$. Putting (8c) into (8a) one obtains a relation between $A'$ and $A$.

$$A' = \frac{1}{A} \tag{9}$$

From (8), (9) one finds as expected that the squared 4-momentum is not invariant under deformed Lorentztransformations.

$$p'^2 = A^2 \, p^2 \tag{10}$$

Requiring that the deformed dispersion relation (7) must be invariant under the transformations in (8) one obtains the transformation of the deformation function $F$

$$F' = \frac{F}{A} \tag{11}$$

The relations (9),(10),(11) can be verified in the models [8],[9].

## 3. Transformation of derivatives and deformed coordinates

At first we transform the SR derivatives $\frac{\partial}{\partial \pi^\mu}$ to DSR variables $p^\mu$. Applying the derivative $\frac{\partial}{\partial \pi^\mu} = \frac{\partial p^\nu}{\partial \pi^\mu} \frac{\partial}{\partial p^\nu}$ together with (6b) to $F^{-1}(\pi) = 1/F(p)$, solving for $\frac{\partial F^{-1}}{\partial \pi^\mu}$ and inserting back one finds with $F_\mu = \frac{\partial F}{\partial p^\mu}$ for the derivatives expressed through nonlinear momenta $p_\mu$ as in [9]

$$\frac{\partial}{\partial \pi^\mu} = \frac{1}{F}\left(\frac{\partial}{\partial p^\mu} - b_\mu \, p^\lambda \, \frac{\partial}{\partial p^\lambda}\right), \; b_\mu = \frac{F_\mu}{F + p^\lambda F_\lambda} \tag{12}$$

In the model [8] we have $b_\mu = \ell \, \delta_{\mu 0}$, while in the model [9] $b_\mu = \ell^2 \, p_i \, \delta_{\mu i}$. The SR coordinates can thereby be expressed in DSR variables $\xi_\mu = i \frac{\partial}{\partial \pi^\mu} = \frac{i}{F}\left(\frac{\partial}{\partial p^\mu} - b_\mu \, p^\lambda \, \frac{\partial}{\partial p^\lambda}\right)$ and transform as first order tensor under standard Lorentz transformations according to (5).

Following the same steps as before (12), we consider the deformed Lorentz transformation of derivatives with respect to DSR momenta $p^\mu$ together with (8) and (9)

$$\frac{\partial}{\partial p'^\mu} = \frac{\partial p^\nu}{\partial p'^\mu} \frac{\partial}{\partial p^\nu} = \left(\frac{1}{A}\overline{\Lambda}_\mu^{\;\nu} + A \, p^\nu \, \frac{\partial A'}{\partial p'^\mu}\right) \frac{\partial}{\partial p^\nu} \tag{13}$$

We apply this derivative to $A'$, use on the right side equation (9) and solve for $\frac{\partial A'}{\partial p'^\mu}$. Substituting back in (13) finally gives the desired transformation of the derivatives under deformed Lorentz transformations where $A_\nu = \frac{\partial A}{\partial p^\nu}$.

$$\frac{\partial}{\partial p'^\mu} = \frac{1}{A}\overline{\Lambda}_\mu^{\;\nu}\left(\frac{\partial}{\partial p^\nu} - a_\nu \, p^\lambda \, \frac{\partial}{\partial p^\lambda}\right), \; a_\nu = \frac{A_\nu}{A + p^\lambda A_\lambda} \tag{14}$$

Again the $a_\nu$ can be calculated in the models [8] as $a_0 = \ell \, (1 - \gamma)$, $a_1 = \ell \, \gamma \, v$ and [9] as $a_0 = \ell^2 \, \gamma^2 \, v \, (p_1 - v \, E)$, $a_1 = \ell^2 \, \gamma^2 \, v \, (E - v \, p_1)$ where $a_2 = a_3 = 0$ in both cases. From (14) one sees that the DSR derivatives transform inhomogenously under deformed Lorentz transformations, which makes it unlikely that they can be used for coordinates similar to SR. Taking instead $x_\mu = \xi_\mu$ would yield commutative coordinates transforming as first order tensors under standard Lorentz transformations, which ditto seems not very attractive.

To identify the coordinates, we recall from [1] that the deformed boost and rotation generators $M_{\mu\nu}$ can obtained from the undeformed ones simply by transforming the SR variables to DSR variables. This is equivalent to the demand that the boost and rotation generators are invariant under this transformation identical to the U-map.

$$\tilde{M}_{\mu\nu} = \pi_\nu \, \xi_\mu - \pi_\mu \, \xi_\nu = p_\nu \, x_\mu - p_\mu \, x_\nu = M_{\mu\nu} \tag{15}$$

Substituting now (3) and (6) together with (12) in this equation, one can easily read off the momentum space representation of the DSR coordinates as $x_\mu = F \, \xi_\mu$ or

$$x_\mu = i \left(\frac{\partial}{\partial p^\mu} - b_\mu \, p^\lambda \, \frac{\partial}{\partial p^\lambda}\right) \tag{16}$$

The transformation of these coordinates in momentum space can be obtained from the transformation of the $\xi_\mu$ from (5) together with (11) or from the transformation of (15) as

$$x'_\mu = \frac{1}{A} \bar{\Lambda}_\mu^{\ \nu} x_\nu \ , \ x'^\mu = \frac{1}{A} \Lambda^\mu_{\ \nu} x^\nu \tag{17}$$

In [4] (second approach) it was recently suggested to fix position space from momentum space by the condition demanding that $p_\mu x^\mu$ remains invariant under the U-map connecting DSR and SR variables allowing for plane wave solutions.

$$\pi_\mu \xi^\mu = p_\mu x^\mu \tag{18}$$

Replacing $\pi_\mu = F p_\mu$ gives immediately $x^\mu = F \xi^\mu$ equivalent to (16). The transformation of the coordinates in [4],[5] or from $p'_\mu x'^\mu = p_\mu x^\mu$ agrees with (17). It is certainly surprising, that coordinates transform differently from the momenta contrary to to SR case. The other option would be coordinates transforming as the momenta i.e. $x_\mu = \frac{1}{F} \xi_\mu$ also leading to a noncommutative space-time, however at the cost of a deformed Lorentz algebra.

From (8) and (14) one receives the transformation of $p'^\lambda \frac{\partial}{\partial p'^\lambda}$ and then from (17) the transformation of $b_\mu$ under deformed Lorentz transformations.

$$p'^\lambda \frac{\partial}{\partial p'^\lambda} = (1 - p^\nu a_\nu) p^\lambda \frac{\partial}{\partial p^\lambda} \ , \ b'_\mu = \frac{1}{A(1-p^\lambda a_\lambda)} \bar{\Lambda}_\mu^{\ \nu} (b_\nu - a_\nu) \tag{19}$$

## 4. Deformed Poincare Heisenberg algebra

The deformed boost and rotation generators are derived from (15) together with (12) and (6) yielding the same expression as in [9]

$$M_{\mu\nu} = i \left( p_\nu \frac{\partial}{\partial p^\mu} - p_\mu \frac{\partial}{\partial p^\nu} + B_{\mu\nu} p^\lambda \frac{\partial}{\partial p^\lambda} \right), \ B_{\mu\nu} = p_\mu b_\nu - p_\nu b_\mu \tag{20}$$

Under deformed Lorentz transformations they behave as $M'_{\mu\nu} = \bar{\Lambda}_\mu^{\ \rho} \bar{\Lambda}_\nu^{\ \sigma} M_{\rho\sigma}$. The generators commute with $F^2 p^2$ but not with $p^2$. The coefficientes $B_{\mu\nu}$ are determined as $B_{0i} = -\ell p_i$ in [8] and $B_{0i} = -\ell^2 E p_i$ in [9] with $B_{ij} = 0$ in both models.

From the momentum space representation of coordinates (16) and boost and rotation generators (20) one derives the deformed algebra replacing the SR Poincare Heisenberg algebra.

$$\begin{aligned}
&[p_\mu, p_\nu] = 0 \\
&[x_\mu, x_\nu] = i(b_\mu x_\nu - b_\nu x_\mu) \\
&[x_\mu, p_\nu] = i(\eta_{\mu\nu} - b_\mu p_\nu) \\
&[M_{\mu\nu}, x_\sigma] = i(\eta_{\mu\sigma} x_\nu - \eta_{\nu\sigma} x_\mu - B_{\mu\nu} x_\sigma) \\
&[M_{\mu\nu}, p_\sigma] = i(\eta_{\mu\sigma} p_\nu - \eta_{\nu\sigma} p_\mu + B_{\mu\nu} p_\sigma) \\
&[M_{\mu\nu}, M_{\rho\sigma}] = i(\eta_{\mu\rho} M_{\nu\sigma} - \eta_{\nu\rho} M_{\mu\sigma} - \eta_{\mu\sigma} M_{\nu\rho} + \eta_{\nu\sigma} M_{\mu\rho})
\end{aligned} \tag{21}$$

(21a) says that momenta remain commutative, so they can be considered as "good" variables in the high energy regime. According to (21b) the coordinates do not commute and therefore must be considered as "bad" variables. They show the Lie-algebra noncommutativity discussed in [10]. From (21c) one sees that the modified commutators between coordinates and momenta give a generalized Heisenberg uncertainty principle. (21d,e) are the deformed commutators of coordinates and momenta with the boost and rotation generators. Finally from (21f) one sees that the Lorentz algebra remains undeformed.

The above relations can be applied to model [8] using $b_\mu = \ell \delta_{\mu 0}$ and $B_{0i} = -\ell p_i$. (21b,c,d) result in the following nonvanishing commutators, where $N_i = M_{0i}$

$$\begin{aligned}
&[x_0, x_i] = i\ell x_i \ , \ [x_i, p_j] = i\delta_{ij} \ , \ [x_0, p_i] = -i\ell p_i \ , \ [x_0, E] = i(1 - \ell E) \\
&[N_i, x_0] = i(-x_i + \ell p_i x_0) \ , \ [N_i, x_j] = i(-\delta_{ij} x_0 + \ell p_i x_j)
\end{aligned} \tag{22}$$

which are in agreement with the Poisson brackets c.f. equations (16),(17) in [5].

In the model [9] we have $b_\mu = \ell^2 p_i \delta_{\mu i}$ and $B_{0i} = -\ell^2 E p_i$ giving the commutators

$$[x_i, x_j] = i\ell^2 (p_i x_j - p_j x_i), \quad [x_0, x_i] = -i\ell^2 p_i x_0$$
$$[x_0, E] = i, \quad [x_i, p_j] = i(\delta_{ij} - \ell^2 p_i p_j) \qquad (23)$$
$$[N_i, x_0] = i(-x_i + \ell^2 E p_i x_0), \quad [N_i, x_j] = i(-\delta_{ij} x_0 + \ell^2 E p_i x_j)$$

Both spatial and temporal coordinates do not commute and don't have sharp values together in this model. Interestingly the fourth relation in (23) for $i = j = 1$ would result in the standard minimum length uncertainty relation in the case $\ell^2 \to -\ell^2$.

## 5. Summary

In summary we have introduced space-time coordinates in DSR theories with energy independent speed of light demanding that the boost and rotation generators are invariant under the U-map in agreement with the recent suggestion of the invariance of $p_\mu x^\mu$ in [4]. Thereby coordinates transform differently from the momenta under deformed Lorentz transformations. As main result we obtain the momentum space representation of coordinates, the deformed algebra and a noncommutative space-time. One may generalise the present ansatz to variable speed of light models [11] with two different functions in the deformed dispersion relation (7) and since they constitute a generalisation one would expect that they also show a space-time noncommutativity.

Once it is known, that the space-time in DSR theories becomes noncommutative with a Lie-algebra structure [10], one can adress questions of backtransformations to coordinate space, noncommutative field theory etc.